# Automated Analysis of Femoral Artery Calcification Using Machine Learning Techniques

Liang Zhao
Computer Science & Engineering
U. of South Carolina
Columbia, SC, USA
lz4@email.sc.edu

Brendan Odigwe
Computer Science & Engineering
U. of South Carolina
Columbia, SC, USA
Bodigwe@email.sc.edu

Susan Lessner
Cell Biology & Anatomy
U. of South Carolina School of Medicine
Columbia, SC, USA
Susan.Lessner@uscmed.sc.edu

Daniel G. Clair
Surgery
U. of South Carolina School of Medicine
Columbia, SC, USA
Daniel.Clair@uscmed.sc.edu

Firas Mussa
Zucker School of Medicine
New Hyde Park, NY, USA
fmussa@northwell.edu

Homayoun Valafar
Computer Science & Engineering
U. of South Carolina
Columbia, SC, USA
homayoun@cse.sc.edu

*Abstract* — We report an object tracking algorithm that combines geometrical constraints, thresholding, and motion detection for tracking of the descending aorta and the network of major arteries that branch from the aorta including the iliac and femoral arteries. Using our automated identification and analysis, arterial system was identified with more than 85% success when compared to human annotation. Furthermore, the reported automated system is capable of producing a stenosis profile, and a calcification score similar to the Agatston score. The use of stenosis and calcification profiles will lead to the development of better-informed diagnostic and prognostic tools.

## I. INTRODUCTION

Peripheral arterial disease (PAD), which results from atherosclerotic plaque buildup with or without calcification in the large arteries of the extremities, constitutes a growing medical burden to the aging population in the United States. Estimated prevalence of PAD increases dramatically with age, from 0.9% in the population 40-49 years of age to 14.5% in people older than 69.[1] In 2001, estimated cost to the US Medicare program for PAD-related treatment was greater than $4.3 billion.[2] This figure does not take into account lost wages and productivity in PAD patients as a result of decreased mobility. Clinical presentation and outcomes of PAD are highly variable, ranging from asymptomatic disease to intermittent claudication (IC, limb pain during exercise) or rest pain in the affected limb. In the most severe cases, chronic limb threatening ischemia (CLTI) may result in tissue loss or gangrene and the need for amputation. Vascular calcification, which is commonly observed in PAD patients with co-morbid diabetes or chronic kidney disease (CKD), complicates interventional treatment and correlates with increased morbidity and mortality.[3]–[5] Currently, there is no validated, widely accepted metric to quantify the extent of arterial calcification in the lower extremities,[5] that would be comparable to the widely accepted Agatston score for coronary artery calcification.[6] Thus, there is an urgent need for better methods to predict the progression of PAD and to identify patients at greatest risk for the most severe clinical outcomes, in order to treat these patients more aggressively. To that end we have begun an exploration in the use of Machine Learning techniques to automate the diagnostic and prognostic processes in relation to PAD.

Artificial intelligent (AI) based predictive tools such as Artificial Neural Networks (ANNs)[7]–[9] have been implemented in recommender systems to enhance shopping experience (Netflix and Amazon), used to assist with speech to text recognition[10] on cell phones, and incorporated into automated driving systems[11] (Uber and Google self-driving vehicles). Extensive work has been performed to create new mechanisms of predictive modeling[12] (such as LSTMs, DNNs, and CNNs) as well as to improve already existing methods (such as ANNs, decision trees, etc). In the closer domain of healthcare[9], [13], ANNs have been used in medical fraud detection[14], [15] and in prediction of a patient's response to a drug[16], [17], thereby paving the way to personalized medicine. More specific to our investigation, the key question for clinicians, patients, regulators and insurance providers is: Can Artificial Intelligence use personalized data in patients with PAD to derive a predictive model of individual outcomes following femoral endarterectomy in terms of complications (death, lower extremity amputation, disease progression and need for intervention) at presentation?

While ANNs have advanced substantially over the last few years, their impact and integration in the field of medical diagnostics has been minimal. The lack of integration of Machine Learning techniques in Clinical Decision Systems and medical science is due to the challenging nature of analyzing medical data, and the absence of well annotated and relevant data appropriate for use in Machine Learning endeavors. In this report, we

This work was funded by NIH grant number P20 RR-016461 to Dr. Valafar and HL145064-01 to Dr. Lessner have investigated both of these challenges and present our intermediate progress with the potential of mitigating some of the long-standing challenges. More specifically, we have developed an unsupervised mechanism of tracking the aorta in X-ray computed tomography angiogram (CTA) scans in a small cohort of 5 patients. Our object-tracking mechanism has successfully tracked the aorta and its branches descending as far down as the patellar surface of the femur with more than 85% success. Using this information, we have created a clustering mechanism of identifying arterial wall, lumen, and the total blockage at each slice of the CTA scan. Finally, we discuss a mixed method approach to creating annotated data appropriate for the task of supervised learning while minimizing the time-requirement of the field-experts.

## II. BACKGROUND AND METHOD

**Previous and Related Work**

Image segmentation is the process of identifying one or multiple objects in an image. Multiple approaches to image segmentation have been proposed and investigated over the years including thresholding[18], clustering[19], motion[20], edge detection[21], or parametric[22] methods, to name a few. Applications of image segmentation have been explored in different domains including content-based image retrieval, object detection, face recognition, and medical imaging. More specifically, automated image processing techniques have been used to detect aortic valves using C-arm CTs[23], aortic aneurysms[24], and aortic dissections[25].

**Background Anatomy**

The primary objective of the current work is to track the aorta as it descends in the human body and branches into other smaller arteries. The aorta is the largest artery in the human body, originating at the outlet of the left ventricle and extending anterior to the vertebral column along the abdominal wall before bifurcating into paired common iliac arteries proximal to the pelvis (Figure 1). Major branch arteries extend from the aorta to provide blood to all parts of the systemic circulation. In the abdominal cavity, these branches begin with the celiac artery, just distal to the diaphragm, followed in order by the superior mesenteric, right renal, left renal, and inferior mesenteric arteries. These abdominal branch arteries are directed perpendicular to the longitudinal orientation of the aorta and will largely be ignored for our purposes.

Our interest in this study lies primarily in the vasculature distal to the iliac bifurcation, which provides blood to the lower extremities. Distal to the bifurcation, the common iliac arteries diverge, with each one supplying blood to one leg. The common iliac arteries further divide, giving rise to the internal and external iliac arteries. After passing through the inguinal ligament, the external iliac artery continues as the femoral artery, which further branches to give rise to the deep femoral and superficial femoral arteries.

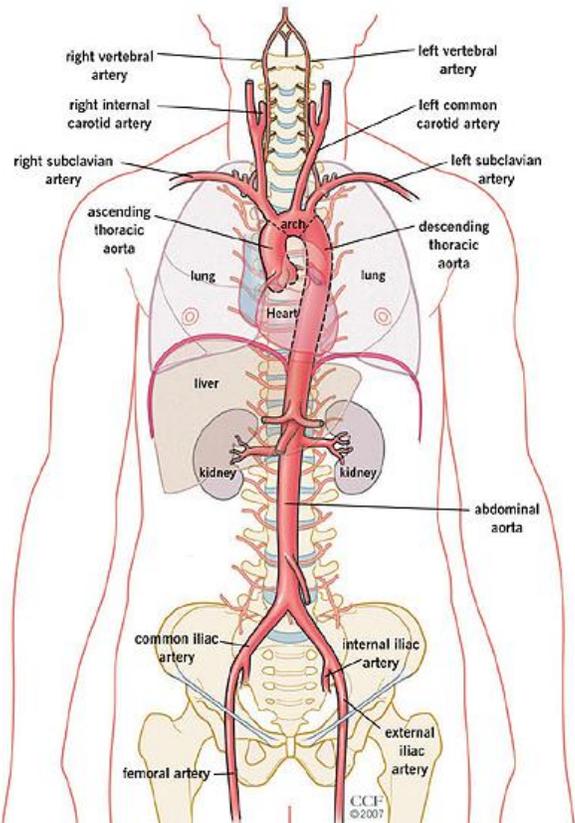

*Figure 1. An illustration of descending aorta and the major arteries including iliac and femoral arteries.*

To better understand the performance of our automated detection of the arterial system starting with the aorta, it is useful to enumerate a simplified number of main arteries that should be expected based on visual inspection of the images. In this study, visual inspections of the data should identify a single aortic artery in the approximate range of transverse slices of 1-100 a brief range of ~20 slices that include two distinguishable arteries (common iliacs), 40 slices with 4 clearly identifiable arteries (internal and external iliac arteries in each leg), that finally branch to numerous smaller arteries.

**Description of Data**

In this study, we analyzed CTA scans with infused contrast agent of five patients who have been diagnosed with peripheral arterial disease. The dataset consisted of both male and female subjects, between 55 and 66 years of age. The CTA scans of these patients were used to perform the automated object tracking approach for identification of the arteries that descend from the aorta. Each image file consisted of a series of approximately 600 transverse images with the resolution of 512x512 that spanned the mid-sternum to the bottom of the subject's feet in the conventional DICOM format. Because of the limited resolution of these images (5mm slices) recognition of arteries below the knee was nearly impossible. Therefore, our region of investigation included the mid-sternum to first appearance of the patella. DICOM files were converted to TIFF files and de-identified prior to analysis.

**Automated Object Tracking and Identification of Calcifications**

The final objective of our investigation is to develop an unsupervised mechanism of annotating the arterial system that starts with the descending aorta. To accomplish this task, we have developed a multistage image segmentation routine using constrained object-tracking that is suitable for this study.

*Stage I – Identification of aorta*: During the first stage of the operation, the first slice of the images is subjected to filtration of pixels based on the min/max thresholds that are provided by the user. In this transverse image, the intensities corresponding to all pixels outside of the allowed thresholds will be set to a value of zero while all pixels with intensities corresponding to the acceptable range will be set to the maximum level. Furthermore, all clusters of pixels that do not adhere to an approximate ovular shape receive an intensity of zero (filtered out).

*Stage II – Tracking of the arterial network:* The primary objective of this stage is to track the descending aorta and the network of arteries that emanate from the aorta. Two primary constraints are used in this stage to identify arteries. The first criterion is based on a cluster of pixels that did not filter during Stage-I and exhibit ovular shape. The second criterion required overlap between any such cluster of pixels and a cluster of pixels identified as an artery from the previous slice.

*Stage III – Identification of Calcification:* After the completion of Stages I and II, the algorithm proceeds with the identification of the calcifications confined to the interior of the identified arterial lumen. The identification of the arterial lumen is a relatively simple task that is accomplished by the clustering of the pixels based on intensity. Pixels with a relatively higher intensity are associated with calcifications.

*Stage-IV: Production of the Final Images:* As the final step in our analysis, all the unobstructed interior portions of the arterial system are marked with blue color, while the calcified regions inside of the arterial lumen are marked as red. Other associated data are produced in CSV file format that summarize the number of vessels, the number of lumen pixels, and the number of calcified pixels for each slice of the images. Such data will be the subject of our future investigations in order to establish the correlative relation between patient outcome and their current CT scans.

**Evaluation Technique**

To evaluate our automated detection mechanism, we will investigate two components. During the first phase of the evaluation, we record the number of vessels identifiable by a human in each slice for each patient. In the second phase, we record the same results for the automated detection of the vessels. Both manual and automated inspections of the images should generally correspond to the general expectations described in section IIB, while anticipating some variations due to natural anatomical variations among people. We also compare the results of the manual annotation of the arteries to that of the automated identification mechanism. Finally, we utilize the summary information reported for each slice of the images, to generate various visual analytics for each patient.

## III. RESULTS AND DISCUSSION

### A. Automated Identification of the Arterial Network

The automated detection process starts with the detection of the aorta in the first slice of the images. Figure 2a illustrates the first slice for patient 6572. After completing Stage-I of our analysis the masking binary image shown in Figure 2b is identified. This mask was then used to track the descending aorta (show in Figure 2c) while identifying the internal calcifications (Figure 3).

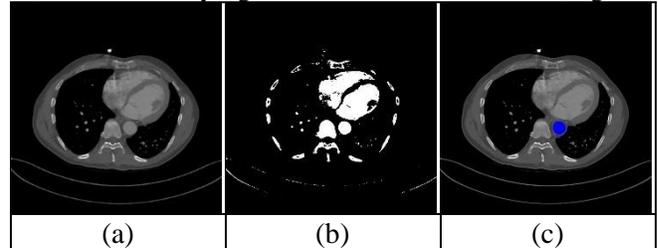

*Figure 2. The first slice for patient 6572 (a) at the beginning of the process, (b) binarized based on cutoff thresholds, and (c) after the completion of the Stage-III.*

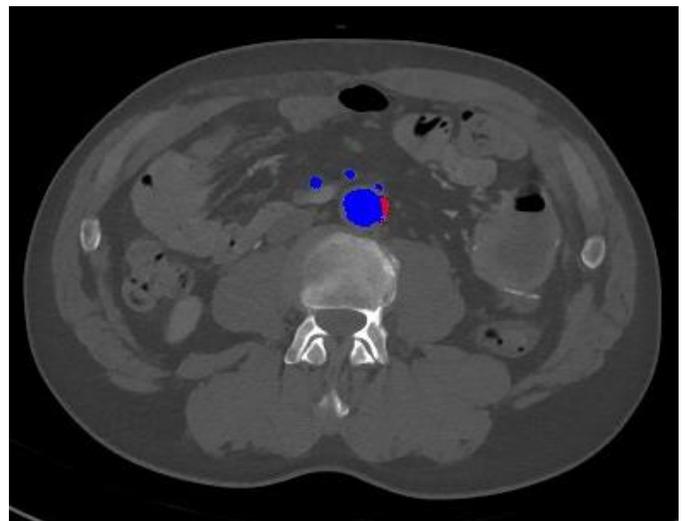

*Figure 3. Slice #84 for patient 6572 at the end of Stage-IV of the automated analysis. Blue indicates the unobstructed lumen of vessels, while red denotes the identified calcification.*

## B. Comparison of Automated and Manual Tracking of Vessels

To validate the outcome of our automated tracking of the descending aorta and the emanating network of arteries, we compare the results to manual inspection and identification of arteries. This comparison can be performed at various levels of granularity spanning from coarse total number of arteries detected for each person, to the fine-grained pixel level. In this investigation, we focus on an intermediate level that compares the number of arteries in each slice detected by the automated system versus manually by a trained human. Figure 4 illustrates the results for manual and automated detection of vessels in slices 1-385. Note the degree of similarity between the two approaches. Furthermore, it is noteworthy that not only in some instances the automated approach has missed detection of vessels (usually slices 250-300), however in slices 80-120, the automated method has over-detected the number of vessels.

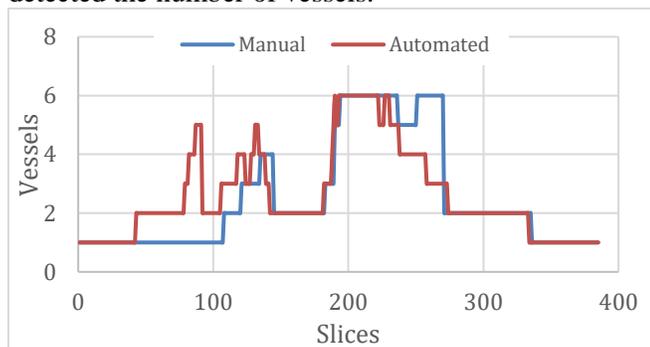

*Figure 4. Comparison of the number of vessels detected manually (blue) to automated (red) for slices 1-385 in patient 6572.*

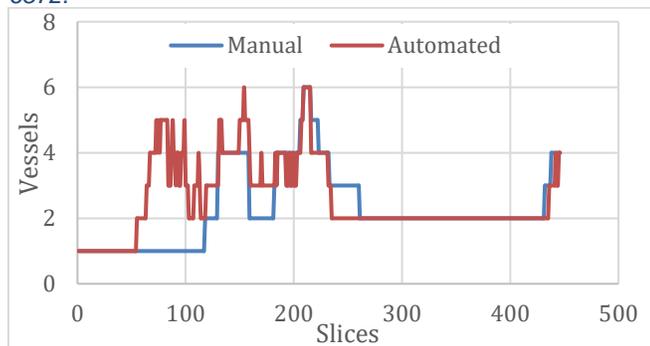

*Figure 5. Comparison of the number of vessels detected manually (blue) to automated (red) for slices 1-385 in patient 6573.*

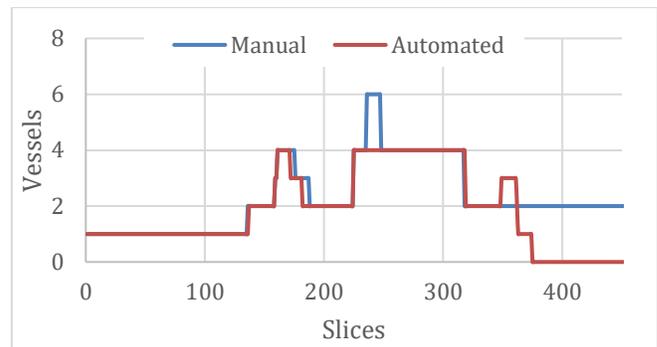

*Figure 6. Comparison of the number of vessels detected manually (blue) to automated (red) for slices 1-385 in patient 6574.*

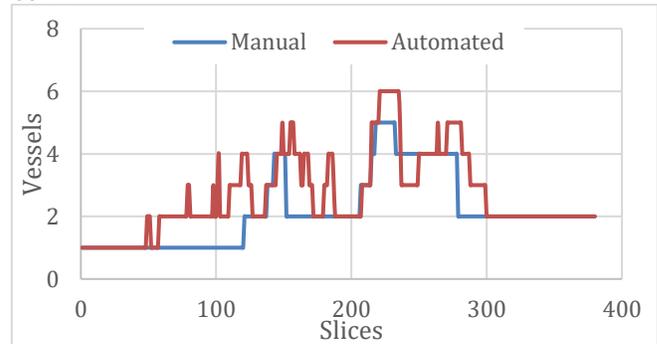

*Figure 7. Comparison of the number of vessels detected manually (blue) to automated (red) for slices 1-385 in patient 6575.*

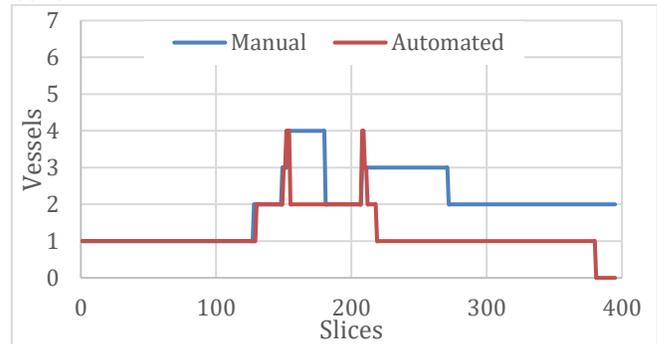

*Figure 8. Comparison of the number of vessels detected manually (blue) to automated (red) for slices 1-385 in patient 6576.*

**Stenosis Profile**

Subsequent to the detection of the arterial system, our automated analysis can discriminate between the lumen or a calcified region of a vessel based on the intensity of the pixels. Using this method, we can quantify several relevant parameters including the volume of the cross-section of a vessel, relative unobstructed area, and the relative area of the obstruction. Using this information, a profile of vessel stenosis for a given patient can be constructed. Figure 9 illustrates the automatically generated stenosis profile for patient 6572 that indicates the highest level of blockage (20%) at slice range of 96-98. In contrast to this patient, the stenosis profile for patient 6573 exhibits a more serious vascular disease with nearly 100% blockage at slice range of 400-408.

In addition to the volumetric information that is purely based on the number of pixels, our automated system can provide a measure of blockage that includes a degree of calcification. Although not discussed here, this score is closely correlated to the Agatston score[6] and is based on the integral of intensities of pixels that are identified as calcifications.

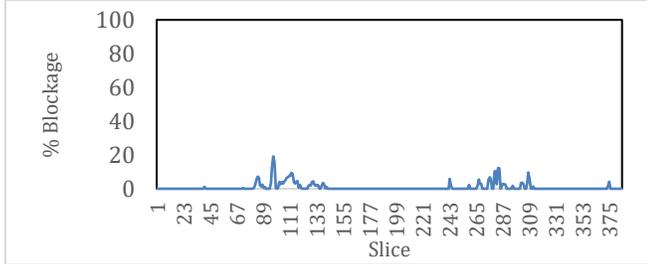
*Figure 9. Automated calculation of the stenosis profile for patient 6572.*

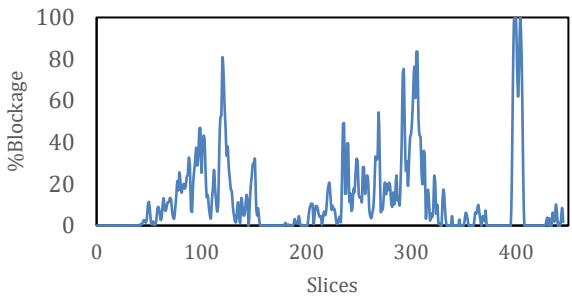
*Figure 10. Automated calculation of the stenosis profile for patient 6575.*

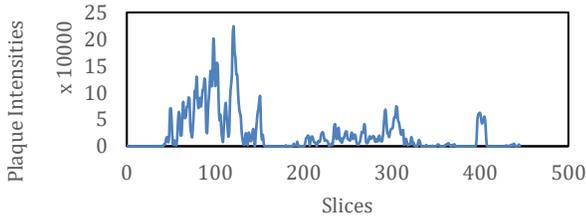
*Figure 11. Sum of the calcification intensities in each slice for patient 6575.*

**Future Work**

Our ultimate goal in this work is to automate identification and quantitation of vessel calcification in CT angiograms of the extremities in patients diagnosed with peripheral arterial disease as a prognostic tool for risk stratification. To this end, we have initiated clinical studies to track patient outcomes and relate adverse outcomes (disease progression, need for additional intervention, amputation, or death) to the extent and distribution of calcification at initial CTA evaluation. Our near-future objective is to train and test more sophisticated ANN based approaches using the annotated images from our object tracking approach.

## IV. CONCLUSION

Detailed annotation of medical images at the pixel level is one pragmatic limitation in initiating multidisciplinary research between experts in the fields of Machine Learning and medical sciences. Here we have presented a mixed-approach that combines thresholding, motion detection, and other geometrical constraints in order to facilitate an effective means of tracking the descending aorta, as it expands into a network of arteries in the lower abdominal and femoral regions. Our object tracking approach has exhibited more than 85% agreement with human annotation of images. Such an approach can be used to provide a more efficient means of image annotation by experts and physicians who cannot dedicate as much time to this kind of activity.

In addition to annotation of the arteries, we have also demonstrated the feasibility of detection and quantification of the calcifications in the major arteries extending into the femoral regions. Tools such as the stenosis profile and Agatston scoring profile can be useful in diagnosis and to correlate with patient outcomes for the eventual development of prognostic tools.